\documentstyle[12pt]{article}
\catcode`\@=11 \@addtoreset{equation}{section}\catcode`\@=12
\newcommand{\beq}[1]{\begin{equation}\label{#1}}
\newcommand{\eeq}{\end{equation}}
\newcommand{\bear}[1]{\begin{eqnarray}\label{#1}}
\newcommand{\ear}{\end{eqnarray}}
\newcommand{\nn}{\nonumber}

\newcommand{\R}{\mbox{\rm I$\!$R} }
\newcommand{\M}{\mbox{\rm I$\!$M} }
\newcommand{\foom}[1]{\protect\footnotemark[#1]}
\newcommand{\email}[2]{\footnotetext[#1]{e-mail: #2}}%
\newcommand{\Title}[1]{\noindent {\Large #1} \\}
\newcommand{\e}{\mbox{\rm e}}
\newcommand{\diag}{\mbox{\rm diag} }
\newcommand{\eps}{\varepsilon }

\begin{document}
\begin{center}
\Title{\large\bf X-fluid and viscous fluid\\
in $D$-dimensional anisotropic integrable cosmology}

\bigskip

\noindent{\normalsize\bf
V. R. Gavrilov\foom 1 and
V. N. Melnikov\foom 2}

\medskip

\noindent{\it Centre for Gravitation and Fundamental Metrology
\\
VNIIMS, 3-1 M. Ulyanovoy St., Moscow 117313, Russia}

\vspace{5mm}
{\bf Abstract}
\end{center}

$D$-dimensional cosmological model describing the evolution of a perfect fluid
with negative pressure (x-fluid) and a fluid possessing both shear and bulk
viscosity in $n$ Ricci-flat spaces is investigated. The second equations
of state are chosen in some special form of metric dependence of the shear
and bulk viscosity coefficients. The equations of motion are integrated and
the dynamical properties of the exact solutions are studied. It is shown
the possibility to resolve the cosmic coincidence problem when the x-fluid
plays role of quintessence and the viscous fluid is used as cold dark
matter.
\email 1 {gavr@rgs.phys.msu.su}
\email 2 {melnikov@rgs.phys.msu.su}

\section{Introduction}
The most probable candidate for the so called quintessence matter
responsible for the current phase of the accelerated expansion of the
Universe is a $\Lambda$ term o more generally an exotic x-fluid -
perfect fluid with negative pressure satisfying a linear barotropic
equation of state (see, for instance, \cite{Waga} and refs. therein).
To describe the present stage of
evolution this x-fluid is to be added to the normal matter, which mainly
consist of cold dark matter. For instance, $\Lambda$CDM model
\cite{Johri},\cite{Bludman} describes the flat Friedman-Robertson-Walker
(FRW) Universe filled with a mixture of quintessence represented by the
cosmological constant $\Lambda$ and  cold dark matter in the form of
pressureless perfect fluid (dust).

However, the flat FRW cosmologies with a x-fluid and a normal perfect
fluid are not free from some difficulties. One of them is the so called
coincidence problem \cite{Steinhardt}: to explain why the quintessence
density and the normal matter density are comparable today, one has to tune
their initial ratio very carefully. The problem may be ameliorated by
replacement the x-fluid by the so called quintessence scalar field -
homogeneous scalar field $Q$ slowly rolling down with some potential $V(Q)$
\cite{Ritis},\cite{Brax}. For instance, the potential
$V(Q)=M^{4+\alpha}Q^{-\alpha},\ \alpha>0$, leads to the so called "tracker"
solution, which "attracts" solutions to the equations of motion before the
present stage for very wide range of initial conditions. Another way for
resolving the problem was proposed in \cite{Jakubi}. The idea is to use a
fluid with bulk viscosity in combination with a quintessence scalar field.

The aim of this paper is to show that the cosmic coincidence problem may be
resolved by using a x-fluid as the quintessence and a viscous fluid as the
normal matter. In section 2 we describe the general model and get basic
equations. To integrate the equations of motion we choose the so called
"second equations of state", which provide us with the dependence of the
shear and bulk viscosity coefficients on time,
in some special form of their metric dependence.
The exact solutions for both isotropic and anisotropic case are obtained in
section 3, where their dynamical properties are discussed.

\section{The general model}
We assume the following metric
\beq{2.1}
{\rm d}s^2=-{\rm e}^{2\gamma(t)}{\rm d}t^2  +
\sum_{i=1}^{n}\exp[2x^{i}(t)]{\rm d}s_i^2,
\end{equation}
on the $D$-dimensional space-time manifold
\beq{2.2}
\M=\R\times M_1\times\ldots\times M_n,
\eeq
where ${\rm d}s_i^2$ is the metric of the Ricci-flat factor space $M_{i}$ of
dimension $d_i$, $\gamma(t)$ and $x^{i}(t)$ are scalar functions of the
cosmic time $t$.
$a_i\equiv\exp[x^{i}]$ is the scale factor of the space $M_i$ and
the function  $\gamma(t)$ determines a time gauge.
The synchronous time $t_s$ is defined by the equation
${\rm d}t_s=\exp[\gamma(t)]{\rm d}t$.

We consider a source of gravitational field in the form of 2-component
cosmic fluid. The first one is a perfect fluid with a density
$\rho^{(1)}(t)$ and a pressure $p^{(1)}(t)$. The second component
supposed to be a viscous fluid. It is characterized by a density
$\rho^{(2)}(t)$, a pressure $p^{(2)}(t)$, a bulk viscosity coefficient
$\zeta(t)$ and a shear viscosity coefficient $\eta(t)$. The overall
energy-momentum tensor of the cosmic fluid reads
\beq{2.3}
T_\nu^\mu=\left( \rho^{(1)} + \rho^{(2)} \right)u^\mu u_\nu +
\left( p^{(1)} + p^{(2)}-\zeta\theta\right)
P_\nu^\mu -2\eta\sigma_\nu^\mu,
\eeq
where $u^{\mu}$ is the $D$-dimensional velocity of the fluid,
$\theta=u^\mu_{;\mu}$ denotes the scalar expansion,
$P_\nu^\mu=\delta_\nu^\mu + u^\mu u_\nu $ is the projector on the
$(D-1)$-dimensional space orthogonal to $u^\mu$,
$\sigma_\nu^\mu=
\frac{1}{2}\left(u_{\alpha;\beta}+
u_{\beta;\alpha}\right)P^{\alpha\mu}P^\beta_\nu-
(D-1)^{-1}\theta P^\mu_\nu$ is the traceless shear tensor
and $\mu,\nu=0,1,\ldots,D-1$.

By assuming the comoving observer condition
$u^\mu=\delta_0^\mu\e^{-\gamma(t)}$, the overall energy-momentum tensor may
be written as
\beq{2.4}
T_\nu^\mu=T^{\mu(1)}_\nu + T^{\mu(2)}_\nu,
\eeq
where
\bear{2.5}
\left(T^{\mu(1)}_\nu\right)=
\diag(-\rho^{(1)},p^{(1)},\ldots,p^{(1)}),\\
\label{2.6}
\left(T^{\mu(2)}_\nu\right)=
\diag(-\rho^{(2)},\tilde{p}^{(2)}_1\delta^{k_1}_{l_1},
\ldots,\tilde{p}^{(2)}_n\delta^{k_n}_{l_n}),
\ear
$k_i,l_i=1,\ldots,d_i$ for $i=1,\ldots,n$.
Here $\tilde{p}^{(2)}_i$ denotes the effective pressure including the
dissipative contribution of the viscous fluid in the factor space described
by the manifold $M_i$. It reads
\beq{2.7}
\tilde{p}^{(2)}_i=p^{(2)} - \e^{-\gamma}\left[\zeta\dot{\gamma}_0+
2\eta\left(\dot{x}^i-\frac{\dot{\gamma}_0}{D-1}\right)\right],
\eeq
where
\beq{2.8}
\gamma_0=\sum_{i=1}^nd_ix^i.
\eeq
 Furthermore, we assume that the barotropic equations
of state hold
\begin{equation}\label{2.9}
p^{(\alpha)} =\left(1-h^{(\alpha)}\right)\rho^{(\alpha)},\ \alpha=1,2,
\eeq
where the $h^{(\alpha)]}$ are constants such that
$h^{(1)}\neq h^{(2)}$.

The Einstein equations $R^\mu_\nu-\frac{1}{2}\delta^\mu_\nu R=
\kappa^2T^\mu_\nu$, where $\kappa^2$ is the gravitational
constant, can be written as
$R^\mu_\nu=\kappa^2(T^\mu_\nu-[T/(D-2)]\delta^\mu_\nu)$.
Further, the equations
$R^0_0-\frac{1}{2}\delta^0_0 R=\kappa^2 T^0_0$
and
$R^a_b=\kappa^{2}(T^a_b-[T/(D-2)]\delta^a_b)$, where $a,b=1,\ldots,D$,
give the following equations of motion
\bear{2.10}
\dot{\gamma}_0^2-\sum_{i=1}^n d_i(\dot{x}^i)^2&=&
2\kappa^2\e^{2\gamma}\left(\rho^{(1)} + \rho^{(2)}\right),\\
\label{2.11}
\ddot{x}^i + (\dot{\gamma}_0-\dot{\gamma})\dot{x}^i&=&
\kappa^2\e^{\gamma}
\left[
\frac{\e^{\gamma}}{D-2}\sum_{\alpha=1}^2h^{(\alpha)}\rho^{(\alpha)} +
\frac{\zeta}{D-2}\dot{\gamma}_0-
2\eta\left(\dot{x}^i-\frac{\dot{\gamma}_0}{D-1}\right)
\right],
\ear
$i=1,\ldots,n$.

The energy conservation law
$\bigtriangledown_\mu T_0^{\mu(2)}=0$ for a viscous fluid described by a
tensor given by equation (\ref{2.6}) reads
\beq{2.12}
\dot{\rho}^{(2)}+\sum_{i=1}^n d_i\dot{x}^i
\left(\rho^{(2)}+\tilde{p}^{(2)}_{i}\right)=0.
\eeq
Owing to the constraint $\bigtriangledown_\mu T_0^\mu=0$ for the overall
energy-momentum tensor given by equation (\ref{2.3}) the similar energy
conservation law is valid for the perfect fluid
\beq{2.13}
\dot{\rho}^{(1)}+
\left(\rho^{(1)}+{p}^{(1)}\right)\sum_{i=1}^n d_i\dot{x}^i=0.
\eeq
Taking into account equation (\ref{2.9}), ones easily integrates
equation (\ref{2.13}). The result is
\beq{2.14}
\rho^{(1)}=Ae^{(h^{(1)}-2)\gamma_0},
\eeq
where $A$ is a positive constant.
We note that the contribution of the perfect fluid component with
$h^{(2)}=2$ to the overall energy momentum tensor is equivalent to the
presence of $\Lambda$-term with $\Lambda=\kappa^2 A$.

Furthermore, by using
equations (\ref{2.10}) and (\ref{2.14})  the presence of the densities
$\rho^{(1)}$ and $\rho^{(2)}$
in equations (\ref{2.11})   can be cancelled. Thus, we obtain the main
governing set of equations
\bear{2.15}
\ddot{x}^i + (\dot{\gamma}_0-\dot{\gamma})\dot{x}^i&=&
\kappa^2 A\frac{h^{(1)}-h^{(2)}}{D-2}
\e^{[h^{(1)}-2]\gamma_0+2\gamma}+
\frac{h^{(2)}}{2(D-2)}
\left(\dot{\gamma}_0^2-\sum_{i=1}^n d_i(\dot{x}^i)^2\right)\nn\\
&+&
\kappa^2\e^{\gamma}
\left[\frac{\zeta}{D-2}\dot{\gamma}_0-
2\eta\left(\dot{x}^i-\frac{\dot{\gamma}_0}{D-1}\right)
\right].
\ear

We use an integration procedure which is based on the $n$-dimensional
Minkowsky-like geometry. Let $\R^n$ be the real vector space and
${\bf e}_1,\ldots,{\bf e}_n$ be the canonical basis in $\R^n$ (i.e.
${\bf e} _1= (1,0,\ldots,0)$ etc). Let us define a
symmetrical bilinear form $\langle,\rangle$ on $\R^{n}$ by
\beq{2.16}
\langle{\bf e}_i,{\bf e}_j\rangle=\delta_{ij}d_j - d_i d_j.
\eeq
Such a form is non-degenerate and has the pseudo-Euclidean signature
$(-,+,\ldots,+)$ \cite{I1} . With this in mind,
a vector ${\bf y}\in\R^n$ is timelike, spacelike or isotropic respectively,
if $\langle{\bf y},{\bf y}\rangle$ takes negative, positive or null
values respectively and two vectors ${\bf y}$ and ${\bf z}$ are orthogonal if
$\langle{\bf y},{\bf z}\rangle=0$.  Hereafter, we use the following vectors
\bear{2.17}
{\bf x}&=&x^1(t){\bf e}_1+\ldots+x^n(t){\bf e}_n,\\
\label{2.18}
{\bf u}&=&u^1{\bf e}_1+\ldots+u^n{\bf e}_n,\ u^i=\frac{-1}{D-2},\ u_i=d_i,
\ear
where the covariant coordinates $u_i$ of the vector ${\bf u}$
are introduced by the usual way. Moreover, we obtain
\beq{2.19}
\langle\dot{\bf x},\dot{\bf x}\rangle=
\sum_{i=1}^n d_i(\dot{x}^i)^2 -\dot{\gamma}_0^2,\
\langle{\bf u},{\bf x}\rangle=\gamma_0,\
\langle{\bf u},{\bf u}\rangle=-\frac{D-1}{D-2}.
\eeq
Thus, using equations (\ref{2.17})-(\ref{2.19}) we rewrite the
main governing set of equations in the following vector form
\bear{2.20}
\ddot{\bf x} + (\dot{\gamma}_0-\dot{\gamma}){\dot{\bf x}}&=&
\left[
\kappa^2 A\left(h^{(2)}-h^{(1)}\right)
\e^{[h^{(1)}-2]\gamma_0+2\gamma}+
\frac{h^{(2)}}{2}\langle\dot{\bf x},\dot{\bf x}\rangle
\right]
{\bf u} \nn\\
&-&
\kappa^2\e^{\gamma}
\left[\left(\zeta+\frac{D-2}{D-1}2\eta\right)\dot{\gamma}_0{\bf u}+
2\eta\dot{\bf x}\right].
\ear
Moreover, the density $\rho^{(2)}$ can be expressed as
\beq{2.21}
\rho^{(2)}=-\frac{\langle\dot{\bf x},\dot{\bf x}\rangle}{2\kappa^2}
\e^{-2\gamma}-\rho^{(1)}.
\eeq

Now we summarize thermodynamics principles. The first law of
thermodynamics applied to the viscous fluid reads
\beq{2.22}
T{\rm d}S={\rm d}\left(\rho^{(2)}V\right) + p^{(2)}{\rm d}V,
\eeq
where  $V$ stands for a fluid volume in the whole space
$M_1\times\ldots\times M_n$,
$S$ is an entropy in the volume $V$ and $T$ is a temperature of the viscous
fluid.  By assuming that the baryon particle number $N_{\rm B}$ in the
volume $V$ is conserved, equation (\ref{2.22}) transforms to
\beq{2.23}
nT\dot{s}=\dot{\rho}^{(2)}+\left(\rho^{(2)}+p^{(2)}\right)
\sum_{i=1}^n d_{i}\dot{x}^i
\eeq
where $s=S/N_{\rm B}$ and $n=N_{\rm B}/V$  stands for the entropy per
baryon and  the baryon number density.
The comparison between
equations (\ref{2.12}) and (\ref{2.23}) gives the variation rate of
entropy per baryon
\beq{2.24}
nT\dot{s}=
\sum_{i=1}^n d_{i}\dot{x}^i\left(p^{(2)}-\tilde{p}^{(2)}_i\right)=
\e^{-\gamma}
\left[
\left(\zeta+\frac{D-2}{D-1}2\eta\right)\dot{\gamma}_0^2+
2\eta\langle\dot{\bf x},\dot{\bf x}\rangle
\right].
\eeq

\section{Exact solutions}
To integrate equation (\ref{2.20}) one needs a second set equations of
state, involving the bulk viscosity coefficient $\zeta$ and the shear
viscosity coefficient $\eta$. Herein, we suppose
\beq{3.1}
\zeta=\frac{\zeta_0}{\kappa^2}\frac{D-2}{D-1}\dot{\gamma}_0\e^{-\gamma},\
\eta=\frac{\eta_0}{2\kappa^2}\dot{\gamma}_0\e^{-\gamma},
\eeq
where $\zeta_0\geq 0$ and $\eta_0$ are constants. When the cosmological model
is isotropic, i.e. $\dot{x}^i=\dot{\gamma}_0/(D-1),\ i=1,\ldots,n$, and the
shear viscosity is not significant, the expression
$H\equiv\dot{\gamma}_0\e^{-\gamma}/(D-1)$  is the Hubble parameter.Then we
get from equation (\ref{3.1}): $\zeta=\frac{D-2}{\kappa^2}\zeta_0H$, i.e.
the bulk viscosity coefficient is linear proportional to the Hubble
parameter. Such kind of the second equation of state describes the so called
"linear dissipative regime" in the FRW world model (see, for instance,
\cite{Jakubi}). As we study an anisotropic cosmological model, we must
involve a shear viscosity as well as a bulk one. So, we propose equations
(\ref{3.1}) for the anisotropic model as a simplest generalization to the
of the second equation of state describing the linear
dissipative regime.

In order to integrate equation (\ref{2.20}), we use the orthogonal basis
\beq{3.2}
\frac{{\bf u}}{\langle {\bf u},{\bf u}\rangle},
{\bf f}_2,\ldots,{\bf f}_n\in\R^n,
\eeq
where the vector ${\bf u}$ was introduced by equation (\ref{2.18})
The orthogonality property reads
\beq{3.3} \langle{\bf u},{\bf f}_j\rangle=0,\ \langle{\bf
f}_j,{\bf f}_k\rangle=\delta_{jk},\quad (j,k=2,\ldots,n).
\eeq
Let us note
that the basis vectors ${\bf f}_2,\ldots,{\bf f}_n$  are space-like, since
they are orthogonal to the time-like vector ${\bf u}$.  The vector ${\bf
x}\in\R^n$ decomposes as follows
\beq{3.4}
{\bf x}=\gamma_0\frac{{\bf
u}}{\langle{\bf u},{\bf u}\rangle} +\sum_{j=2}^n\langle {\bf x},{\bf
f}_j\rangle{\bf f}_j.
\eeq
Hence, under the above assumptions equation
(\ref{2.20}) reads in the terms of coordinates in such basis as follows
\bear{3.5}
&\ddot{\gamma}_0+
\left[
\left( 1-\frac {h^{(2)}} {2} - \zeta_0 \right) \dot{\gamma}_0-\dot{\gamma}
\right]\dot{\gamma}_0&=\nn\\
\frac{D-1}{D-2} &
\left[
\kappa^2 A\left(h^{(1)}-h^{(2)}\right)
\e^{[h^{(1)}-2]\gamma_0+2\gamma}-
\frac{h^{(2)}}{2}\sum_{j=2}^n\langle\dot{\bf x},{\bf f}_j\rangle^2
\right]&,\\
\label{3.6}
&\langle\ddot{\bf x},{\bf f}_j\rangle+
\left[(1+\eta_0)\dot{\gamma}_0-\dot{\gamma}\right]
\langle\dot{\bf x},{\bf f}_j\rangle&=0 \quad (j=2,\ldots,n).
\ear
The integration of equation (\ref{3.6}) gives
\beq{3.7}
\langle\dot{\bf x},{\bf f}_j\rangle=p^j\e^{\gamma-(1+\eta_0)\gamma_0},
\eeq
where $p^j$ is an arbitrary constant.

Further we determine the time gauge by
\beq{3.8}
\gamma=k\gamma_0,
\eeq
where $k$ is a constant.

By substituting the functions $\langle\dot{\bf x},{\bf f}_j\rangle$
into equation (\ref{3.5}) we obtain the following integrable by quadrature
ordinary differential equation
\bear{3.9}
\ddot{\gamma}_0-
\left(k-1+\frac {h^{(2)}} {2}+\zeta_0 \right) \dot{\gamma}_0^2
=\nn\\
\frac{D-2}{D-1}\e^{2(k-1)\gamma_0}
\left[
\kappa^2 A\left(h^{(1)}-h^{(2)}\right)
\e^{h^{(1)}\gamma_0}-
\frac{h^{(2)}}{2}\sum_{j=2}^n
\left(p^j\right)^2\e^{-2\eta_0\gamma_0}
\right].
\ear

In what follows we accept the agreement
\beq{3.10}
{\rm d}t>0,
\eeq
i.e. the cosmic time increases during the evolution.

It should be noted that the solutions of equation (\ref{3.9}) corresponding
to different sets of the parameters $h^{(1)},h^{(2)}, \zeta _0$ and
$\eta_0$ may lead to nonsatisfactory from the physical viewpoint
cosmological evolutions. Further, we study only ones, which satisfy the
following {\it consistency condition}: neither the density $\rho^{(2)}(t)$
nor the variation rate of entropy $\dot{s}(t)$ have negative values on any
time interval.

\subsection{The isotropic model}
The {\em isotropic model\/} is described by the metric given by equation
(\ref{2.1}) with
\beq{3.11}
a\equiv\e^{x^i}=\e^{\gamma_0/(D-1)},\ i=1,\ldots,n.
\eeq
The scale factor $a$ of the whole isotropically evolving space
$M_1\times\ldots\times M_n$ can be obtained by integration of equation
(\ref{3.9}) with
\beq{3.12}
p^j=0,\ j=2,\ldots,n.
\eeq
It can be proven that the above mentioned {\it consistency condition}
leads to the following constraint
\beq{3.13}
h^{(1)}-h^{(2)}-2\zeta_0>0.
\eeq
At first we present the special solution to equation (\ref{3.9}).
The special solution describes the asymptotical behaviour of all solutions
at late time. It is the steady state solution
\beq{3.14}
a\sim\exp\left[\sqrt{\frac{2\Lambda(1+\eps_0)}{(D-1)(D-2)}}t_s\right],\
\rho^{(1)}=A\equiv\Lambda/\kappa^2,
\eeq
for $h^{(2)}=2$ and  shows a power-law behaviour
\beq{3.15}
a\sim t_s^{2/[(2-h^{(1)})(D-1)]},\ \rho^{(1)}\sim t_s^{-2}
\eeq
for $h^{(1)}\neq 2$, where $t_s$ is the synchronous time.
Moreover, we present the deceleration parameter
\beq{3.16}
q\equiv - \frac{a(\ddot{a}-\dot{\gamma}\dot{a})}{\dot{a}^2}
=-1 +\frac{D-1}{2}\left(2-h^{(1)}\right),
\eeq
the density ratio
\beq{3.17}
\rho^{(2)}/\rho^{(1)}=
\eps_0\equiv\frac{2\zeta_0}{h^{(1)}-h^{(2)}-2\zeta_0}
\eeq
and the overall pressure
\beq{3.18}
p^{(1)}+\tilde{p}^{(2)}=\left(1-h^{(1)}\right)(1+\eps_0)\rho^{(1)}.
\eeq
The variation rate of entropy  is positive
(if $\zeta_0>0$) and $nT\dot{s}\sim\rho^{(1)}$.

To obtain the general solution to equation (\ref{3.9}) in the isotropic case
we suppose that the parameter
$k$ specifying the time gauge by equation (\ref{3.8}) is
\beq{3.19}
k=1-h^{(1)}/2.
\eeq
(We note that the time $t$ becomes synchronous if the perfect fluid
component appears as $\Lambda$-term, i.e. $h^{(1)}=2$). This yields that
equation (\ref{3.9}) under the condition given by equation (\ref{3.12})
is integrable by elementary methods. The result is
\bear{3.20} a&=&a_0\left(\frac{\tau^2}{1-\tau^2}\right)
^{1/[(D-1)(h^{(1)}-h^{(2)}-2\zeta_0)]},\\
\label{3.21}
\rho^{(1)}&=&Aa^{(D-1)[h^{(1)}-2]},\\
\label{3.22}
q&=&-1 +\frac{D-1}{2}
\left[
2-h^{(1)}+\left(h^{(1)}-h^{(2)}-2\zeta_0\right)(1-\tau^2)
\right],\\
\label{3.23}
\rho^{(2)}/\rho^{(1)}&=&(1+\eps_0)/\tau^2-1,\\
\label{3.24}
p^{(1)}+\tilde{p}^{(2)}&=&(1+\eps_0)
\left[
\left(1-h^{(2)}-2\zeta_0\right)\frac{1-\tau^2}{\tau^2}+1-h^{(1)}
\right]
\rho^{(1)},
\ear
where we introduced the following variable
\beq{3.25}
\tau=\tanh
\left[
\left(h^{(1)}-h^{(2)}\right)
\sqrt{\frac{\kappa^2(D-1)A}{2(D-2)(1+\eps_0)}}(t-t_0)
\right],\  t>t_0.
\eeq

Now we analyze the role of viscosity in this isotropic 2-component model
using the obtained exact solution. The main features of the model are the
following. Under the weak energy condition
($\rho^{(1)}+\rho^{(2)}\geq 0,\
\rho^{(1)}+\rho^{(2)}+p^{(1)}+\tilde{p}^{(2)}\geq 0$),
which leads to the following restriction on the parameters
$2\geq h^{(1)}>h^{(2)}+2\zeta_0$, the Universe expands eternally from the
initial singularity. Near the singularity we obtain in the main order
$a\sim t_s^{2/[(2-h^{(2)}-2\zeta_0)(D-1)]}$
There is only nonsingular solution given by equation
(\ref{3.14}).

All solutions describes the accelerated expansion at least on the late phase
of evolution if $h^{(1)}>2(D-2)/(D-1)$.
The solutions given by equation (\ref{3.20}) describe the
period of decelerated expansion if
$h^{(2)}+2\zeta_0<2(D-2)/(D-1)$. We note that the cosmic deceleration phase
is important within mechanism of the clumping of matter into galaxies (see,
for instance, \cite{Johri}). Under the assumption
$2\geq h^{(1)}>2(D-2)/(D-1)>h^{(2)}+2\zeta_0$ the decelerated expansion
takes place during the time interval $(t_0,t^*)$, where $t^*$ is defined by
\beq{3.26}
t^*-t_0=
\frac
{{ \rm arcosh}\sqrt{
1+\frac{2(D-2)/(D-1)-h^{(2)}-2\zeta_0}{h^{(1)}-2(D-2)/(D-1)}
}
}
{
\sqrt{
\kappa^2\frac{A(D-1)}{2(D-2)}\left(h^{(1)}-h^{(2)}\right)
\left(h^{(1)}-2\frac{D-2}{D-1}\right)
\left[
1+\frac{2(D-2)/(D-1)-h^{(2)}-2\zeta_0}{h^{(1)}-2(D-2)/(D-1)}
\right]
}
}.
\eeq
Equation (\ref{3.26}) shows that introducing of the bulk viscosity reduces
the phase of the decelerated expansion.

The density ratio given by equations (\ref{3.17}) and (\ref{3.23})
exhibits the following property
\beq{3.27}
{\rm lim}_{t\to +\infty}\rho^{(2)}/\rho^{(1)}=
\eps_0\equiv\frac{2\zeta_0}{h^{(1)}-h^{(2)}-2\zeta_0}.
\eeq
So, the bulk viscosity allow to resolve the coincidence problem which appears
in the unviscous 2-component model.

\subsection{The anisotropic model}
Now we consider the general anisotropic behaviour of the model.
Once integrating equation (\ref{3.9}) we get the first integral
of the form
$\dot{\gamma}_0=F(\gamma_0,C)$, where $F$ is some function and $C$ is an
arbitrary constant. Substituting $\dot{\gamma}_0$ and the functions
$\langle\dot{\bf x},{\bf f}_j\rangle$ given by equation (\ref{3.7})
into equations (\ref{2.21}) and (\ref{2.24}) we express $\rho^{(2)}$
and $nT\dot{s}$ via $\gamma_0$. The subsequent analysis of the expressions
for $\gamma_0$, $\rho^{(2)}$ and $nT\dot{s}$ shows the presence of solutions
with physically nonsatisfactory behaviour near the initial singularity. To
exclude such solutions we put the following restrictions on the parameters
\beq{3.28}
2\geq h^{(1)} > h^{(2)}+2\zeta_0 \geq -2\eta_0 \geq 2\zeta_0,\
-2\eta_0\geq h^{(2)}\geq 0.
\eeq
These restrictions guarantee the following properties for all solutions:
the density $\rho^{(2)}$ and the variation rate of entropy $\dot{s}$ are
positive during the evolution, which starts at the initial singularity of
the Kasner type and proceeds eternally to the subsequent isotropic
expansion;
the density ratio
$\rho^{(2)}/\rho^{(1)}$ tends to the nonzero constant as $t\to +\infty$
(see equation (\ref{3.27})).

To study these behaviour in detail let us obtain the exact solution. We note
that the equation $\dot{\gamma}_0=F(\gamma_0,C)$ is integrable by quadrature
for arbitrary parameters
$h^{(1)},h^{(2)},\zeta_0$ and $\eta_0$. To express the exact solution in
elementary functions we put the following relation on the parameters
\beq{3.29}
h^{(1)}-2h^{(2)}-2\eta_0-4\zeta_0=0.
\eeq
For the nonviscous model ($\eta_0=\zeta_0=0$) the relation reads
$h^{(1)}=2h^{(2)}$. The latter corresponds, for instance, to the so called
$\Lambda$CDM cosmological model with $\Lambda$-term ($h^{(1)}=2$)
and dust ($h^{(2)}=1$). We remind that the parameters obey the inequalities
given by formula (\ref{3.28}). Comparing equation (\ref{3.29}) and formula
(\ref{3.28}), one gets $h^{(2)}>0$.

Now we start the integration procedure in the time gauge defined by
equations  (\ref{3.8}) and (\ref{3.19}) . The integration of equation
(\ref{3.9}) gives
\beq{3.30}
\e^{\beta\gamma_0(t)}=
C_0\frac{\tau(\sin\alpha+\tau\cos\alpha)}{1-\tau^2},
\eeq
where we cancelled the constant $\sum_{j=2}^n(p^j)^2$ by introducing the
following constants
\beq{3.31}
C_0\geq
\sqrt{\frac{2h^{(2)}}{\kappa^2A(h^{(1)} - h^{(2)})}\sum_{j=2}^n(p^j)^2},\
\alpha=\arcsin\frac{
\sqrt{\frac{2h^{(2)}}{\kappa^2A(h^{(1)} - h^{(2)})}\sum_{j=2}^n(p^j)^2}}
{C_0}
\in[0,\pi/2],
\eeq
the parameter $\beta$ is defined as follows
\beq{3.32}
\beta=h^{(1)} - h^{(2)} - 2\zeta_0>0.
\eeq
The variable $\tau$ was introduced by equation (\ref{3.25}).
Substituting equation (\ref{3.30}) into equation (\ref{3.7}) and
taking into account equations (\ref{3.8}) and (\ref{3.19}),
we obtain by integration
\beq{3.33}
\langle {\bf x},{\bf f}_j\rangle=
\sqrt{\frac{(D-2)\beta}{(D-1)h^{(2)}\sum_{j=2}^n(p^j)^2}}p^j
\ln\left[\frac{\tau}{\sin\alpha+\tau\cos\alpha}\right]^{1/\beta}+q^j,
\eeq
where $q^j$ are arbitrary constants. Substituting
equations (\ref{3.30}) and (\ref{3.33}) into the decomposition given by
equation (\ref{3.4}), we get
\beq{3.34}
{\bf x}=
\ln\left[C_0\frac{(\sin\alpha+\tau\cos\alpha)^2}{1-\tau^2}\right]
^{1/\beta}
\frac{{\bf u}}{\langle{\bf u},{\bf u}\rangle}+
\ln\left[ \frac{\tau}{\tau(\sin\alpha+\tau\cos\alpha)} \right ]
^{1/\beta}
{\bf r} + {\bf s},
\eeq
where the vectors ${\bf r}\in\R^n$ and ${\bf s}\in\R^n$ are defined
as follows
\beq{3.35}
{\bf r}\equiv\sum_{i=1}^nr^i{\bf e}_i=
\sqrt{\frac{(D-2)\beta}{(D-1)h^{(2)}\sum_{j=2}^n(p^j)^2}}
\sum_{j=2}^n p^j{\bf f}_j+
\frac{{\bf u}}{\langle{\bf u},{\bf u}\rangle},\
{\bf s}\equiv\sum_{i=1}^ns^i{\bf e}_i=
\sum_{j=2}^n q^j{\bf f}_j.
\eeq
Owing to the orthogonality property given by equation (\ref{3.3})
the coordinates $r^i$ and $s^i$ of these vectors in the canonical
basis ${\bf e}_1,\ldots,{\bf e}_n$ satisfy the following constraints
\bear{3.36}
\sum_{i=1}^nd_ir^i=\langle{\bf u},{\bf r}\rangle=1,\
\sum_{i=1}^nd_i\left(r^i\right)^2=
\langle{\bf r},{\bf r}\rangle-
\sum_{i,j=1}^nd_id_jr^ir^j=\frac{(D-2)\beta}{(D-1)h^{(2)}}
+\frac{1}{D-1},\\
\label{3.37}
\sum_{i=1}^nd_is^i=\langle{\bf u},{\bf s}\rangle=0.
\ear
The constants $r^i$ may be called Kasner-like parameters, because of the
existence of the constraints given by equation (\ref{3.36}). We remind, that
the coordinates of the vector ${\rm x}$ in the canonical basis are the
logarithms of the scale factors $a_i\equiv\exp[x^{i}]$.

Finally, we present the exact solution:
\bear{3.38}
{\rm d}s^2&=&-
\left[
C_0\frac{\tau(\sin\alpha+\tau\cos\alpha)}{1-\tau^2}
\right]
^{(2-h^{(2)})/\beta} {\rm d}t^2 +
\left[C_0\frac{(\sin\alpha+\tau\cos\alpha)^2}{1-\tau^2}\right]
^{2/(D-1)\beta}\times\nn\\
&\times&
\sum_{i=1}^n\left[ \frac{\tau}{\sin\alpha+\tau\cos\alpha} \right ]
^{2r^i/\beta}\e^{2s^i}{\rm d}s_i^2,
\ear
\bear{3.39}
\rho^{(1)}=A\left[
C_0\frac{\tau(\sin\alpha+\tau\cos\alpha)}{1-\tau^2}
\right]
^{-(2-h^{(2)})/\beta},
\ear
\bear{3.40}
\rho^{(2)}/\rho^{(1)}=(1+\eps_0)
F^2_1(\tau)\left[1-\frac{\beta}{h^{(2)}}F^2_2(\tau)\right]-1,
\ear
\bear{3.41}
nT\dot{s}=
\sqrt{8\kappa^2A\frac{D-1}{D-2}(1+\eps_o)^3}
F^3_1(\tau)
\left[\zeta_0+\frac{\eta_0\beta}{h^{(2)}}F^2_2(\tau)\right]\rho^{(1)},
\ear
where
\beq{3.42}
F_1(\tau)=\frac
{\frac{1}{2}(1+\tau^2)\sin\alpha+\tau\cos\alpha}
{\tau(\sin\alpha+\tau\cos\alpha)},\
F_2(\tau)=\frac
{(1-\tau^2)\sin\alpha}
{(1+\tau^2)\sin\alpha+2\tau\cos\alpha}.
\eeq
The solution has the following integration constants:
$C_0>0$, $\alpha\in[0,\pi/2]$, $t_0$, $r^1,\ldots,r^n$, $s^1,\ldots,s^n$.
The constants $r^i$ and $s^i$ satisfy the constraints given by equations
(\ref{3.36}) and (\ref{3.37}). Then the number of free integration constants
is $2n$ as required. The limit for $\alpha\to+0$ of this exact solution is
the isotropic solution obtained in section 3.2.

Before we start the studying of the obtained exact solution
near the initial singularity
let us
remind the multi-dimensional generalization of the well-known {\em Kasner
solution\/} \cite{Ivas}. It reads (for the synchronous time
$t_s$) as follows
\beq{3.43} {\rm d}s^2=-{\rm
d}t^2_s+\sum_{i=1}^{n}A_it_s^{2\eps^i}{\rm d}s^2_i.
\eeq
Such a metric
describes the evolution of a vacuum model under consideration.  The Kasner
parameters $\eps^i$ satisfy the constraints
\beq{3.44}
\sum_{i=1}^nd_i\eps^i=1,\quad \sum_{i=1}^nd_i\left(\eps^i\right)^2=1.
\eeq
The
generalized Kasner solution describes the contraction of some spaces from
the set $M_1,\ldots,M_n$ and the expansion for the other ones. According to
equation (\ref{3.44}), the number of either contracting or expanding spaces
depends on $n$ (the total number of spaces) and $d_{i=1,n}$ (their
dimensions).

We note that the constraints given by equation (\ref{3.36}) for $r^i$
coincide with these constraints for $\eps^i$ when $\zeta_0=\eta_0=0$
and $h^{(1)}=2h^{(2)}$, i.e. the Kasner-like parameters $r^i$
become exactly Kasner parameters $\eps^i$ in the absence of viscosity.
Therefore, if the parameters $\zeta_0$ and $\eta_0$ are small enough, then
the model describes a behaviour of the Kasner type as $\tau\to+0$, i.e.
towards to the initial singularity. However, too strong viscosity suppress
a behaviour of the Kasner type. It can be shown that if the parameters
$\zeta_0=\eta_0=0$ are large enough, then the model describes expansion of
all factor spaces $M_1,\ldots,M_n$ near the initial singularity.

One can prove that the final stage of the evolution ($\tau\to 1-0$)
exhibits the isotropic expansion. The asymptotical behaviour of the model
for $\tau\to 1-0$ is described by the exact solution given by
equation (\ref{3.14})-(\ref{3.18}).

\vspace{10mm}

{\bf Acknowledgments}
This work was supported  by the Russian Foundation for Basic
Research (Grant 98-02-16414) and DFG Proj. 436 RUS 113/236/0(R).

\pagebreak

\end{document}